\documentclass[final,1p,11pt]{elsarticle}

\usepackage{epsfig}
\usepackage{array,tabularx,epsfig,mathrsfs,graphicx,rotating}
\usepackage{ifthen}
\usepackage{amsfonts}
\usepackage{ragged2e}
\PassOptionsToPackage{hyphens}{url}
\usepackage[hyphens]{url}
\usepackage{hyperref}
\usepackage{listings}
\usepackage{lineno}
\usepackage{subfig}
\usepackage{epstopdf}
\usepackage{color}
\usepackage{float}
\usepackage{verbatim}
\usepackage{color,soul}

\usepackage[normalem]{ulem} % either use this (simple) or
\usepackage{soul} % use this (many fancier options)
\usepackage{amsmath,amssymb}

\let\originallesssim\lesssim
\let\originalgtrsim\gtrsim

\DeclareRobustCommand{\lesssim}{%
  \mathrel{\mathpalette\lowersim\originallesssim}%
}
\DeclareRobustCommand{\gtrsim}{%
  \mathrel{\mathpalette\lowersim\originalgtrsim}%
}

\makeatletter
\newcommand{\lowersim}[2]{%
  \sbox\z@{$#1<$}%
  \raisebox{-\dimexpr\height-\ht\z@}{$\m@th#1#2$}%
}
\makeatother

\hypersetup{
  colorlinks=true,
  linkcolor=blue,
  citecolor=blue,
  urlcolor=blue
}

\graphicspath{{figs/}}

\newcommand{\beq}{\begin{equation}}
\newcommand{\eeq}{\end{equation}}

\chardef\til=126

\journal{ANL-HEP-172661}

\begin{document}
%\hfill ANL-HEP-149528
\definecolor{mygreen}{rgb}{0,0.6,0} \definecolor{mygray}{rgb}{0.5,0.5,0.5} \definecolor{mymauve}{rgb}{0.58,0,0.82}

\lstset{ %
 backgroundcolor=\color{white},   % choose the background color; you must add \usepackage{color} or \usepackage{xcolor}
 basicstyle=\footnotesize,        % the size of the fonts that are used for the code
 breakatwhitespace=false,         % sets if automatic breaks should only happen at whitespace
 breaklines=true,                 % sets automatic line breaking
 captionpos=b,                    % sets the caption-position to bottom
 commentstyle=\color{mygreen},    % comment style
 deletekeywords={...},            % if you want to delete keywords from the given language
 escapeinside={\%*}{*)},          % if you want to add LaTeX within your code
 extendedchars=true,              % lets you use non-ASCII characters; for 8-bits encodings only, does not work with UTF-8
 keepspaces=true,                 % keeps spaces in text, useful for keeping indentation of code (possibly needs columns=flexible)
 frame=tb,
 keywordstyle=\color{blue},       % keyword style
 language=Python,                 % the language of the code
 otherkeywords={*,...},            % if you want to add more keywords to the set
 rulecolor=\color{black},         % if not set, the frame-color may be changed on line-breaks within not-black text (e.g. comments (green here))
 showspaces=false,                % show spaces everywhere adding particular underscores; it overrides 'showstringspaces'
 showstringspaces=false,          % underline spaces within strings only
 showtabs=false,                  % show tabs within strings adding particular underscores
 stepnumber=2,                    % the step between two line-numbers. If it's 1, each line will be numbered
 stringstyle=\color{mymauve},     % string literal style
 tabsize=2,                        % sets default tabsize to 2 spaces
 title=\lstname,                   % show the filename of files included with \lstinputlisting; also try caption instead of title
 numberstyle=\footnotesize,
 basicstyle=\small,
 basewidth={0.5em,0.5em}
}

% \linenumbers

\begin{frontmatter}

\title{
A note on blind technique for new physics searches \\ in particle physics}

\author[add1]{S.V.~Chekanov}
\ead{chekanov@anl.gov}

\address[add1]{
HEP Division, Argonne National Laboratory,
9700 S.~Cass Avenue,
Lemont, IL 60439, USA.
chekanov@anl.gov
}

\begin{abstract}
This paper attempts to classify various blinding strategies
used in particle physics. It argues that the blinding technique
is not used consistently throughout  searches for new physics. More importantly, the blinding technique, in its traditional sense, cannot be applicable for many current and future searches when 
statistical  precision of data  significantly exceeds  the current level of our understanding of Standard Model (SM) backgrounds.
\end{abstract}

\end{frontmatter}

%%%%%%%%%%%%%%%%%%%%%%%%%%%%%%%%%%%%%%%%%%%%%%%%%%%%%%%%%%%%%%%%%%
\section{Introduction}
%%%%%%%%%%%%%%%%%%%%%%%%%%%%%%%%%%%%%%%%%%%%%%%%%%%%%%%%%%%%%%%%%%

A blind analysis is a technique based on 
measurements of event signatures in  ''signal'' regions (i.e. where a signal is expected to show up above some background level) 
using selection cuts developed with the help of theoretical predictions
or data control regions, without looking at signal regions directly 
(see, for example, \cite{Roodman:2003rw,Klein:2005di}). 
The goal of such a technique is to avoid unintended biases that may 
influence  a measurement towards desirable results. On a technical side,
blinding can be applied to shapes of distributions or/and to normalizations of distributions.
% comment from Sam.

Some variations of the blinding technique have been  widely used 
at the Large Hadron Collider (LHC) and other higher-energy particle (HEP)
experiments. Often, this technique is  considered as an official policy in dealing with
preparations of physics analyses for publications. However, published articles  
often lack a proper description of the criteria that define 
the level of rigor of ``blindness'' to signal regions within a broad range of possible ``blinding'' methods.

It is interesting to note that, historically,  no unexpected discoveries beyond the Standard Model (SM) have been made in recent decades using  blinding technique\footnote{It is more difficult to say about how many ``false-positive`` results have been avoided when using this technique since such studies are often did not merit publication, and are usually dismissed after sufficient scrutiny by collaborations.} in its traditional definition (see below).
The observation of the Higgs boson was a special case since its properties were well known prior to its observation, and the existence of the  Higgs boson was expected by many scientists. Its mass was unknown on the theory side, but the experimental limits of previous experiments pointed to the expected mass region for the LHC searches. It is easy to argue that the discovery of the Higgs boson could easily be made even without the blinding method after collecting a sufficient amount of data for SM measurements of invariant-mass distributions (such as $\gamma\gamma$).

% comment from Sam

In this note, we will discuss conceptual limitations of the blinding techniques, and  why the blinding technique in its traditional sense may not be an appropriate method for many searches beyond 
the Standard Model (BSM).  
It is not unreasonable to think that this technique may slow down the pace of discoveries compared to previous 
decades where such technique (in combinations with Monte Carlo simulations) was not widely used. 
Support of this point of view can be  drawn from our analyses of the history of particle physics which will be briefly discussed.

Some specific techniques used for data  blinding  were discussed it \cite{Klein:2005di}.
As correctly pointed out in  \cite{Klein:2005di},  there is no single blinding technique.
Still, we think it is possible to characterize such techniques using broad conceptual terms, 
without giving exact technical details on how the blinding is achieved. 
Below we will attempt to define different classes of blinding procedures used in the past and, more recently,  at the LHC experiments.

\section{Classic case. Type A}

The most classic case of a blinding technique is when theoretical predictions for background and signal distributions are well established beyond the statistical uncertainties expected for  signal regions of data.

Alternatively, theoretical simulations can be replaced by a control region
derived from data.  It  is expected that the control region has statistics as high as the signal region itself, 
has the same physics menu of SM background
processes, and uses the same data reconstruction procedure. 

In this method, an analysis strategy is developed using the 
expected predictions, but hiding the signal region from the analysis teams that develop selection cuts.
Then this strategy is applied to data after ``unblinding''. 
It is expected that several teams (analyzers)  work on developing selection cuts independently and, preferably,  
use independent techniques,  and unblind the signal region at the same time (without biasing conclusions of other teams). 

The results of unblinding should be published independently of actual observations.
No additional manipulations with data are expected prior to publication.
More specifically,  a reduction of discrepancies with the SM, in the case if they are observed, is not allowed.
Examples of such blinding can often be found in particle spectroscopy where
a region of invariant masses is removed while keeping ``side-bands'' of real data.
In high energy physics, the discovery of the Higgs boson is a classical example of the class A blinding \cite{Aad:2012tfa, Chatrchyan_2012}. The  selection procedures were formally
approved and fixed before the results from data in the signal region were examined.
This was possible since reliable predictions for both background and expected signal rates were available, the signal mass region was
known from exclusion limits of previous experiments, and two independent experiments agreed on the strategy for releasing their positive results.

\section{Type B}
Although the  method described above is very straightforward, it should be noted that the 
immediate publication of a discovery by a single experiment (or by a single analysis group) 
is unlikely to occur  without extensive post-unblinding checks.
An application of the  Sagan's standard  "extraordinary claims require extraordinary evidence" implies that it is very unlikely unblinded results from a single experiment  (and, to a more extreme, by a single group within the same experiment) can be published without an extensive evaluation of systematic effects that may cause the unexpected features. All such "post-unblinding" checks do not fall under the ``blinding strategy'' of the type A since data in signal regions can easily be manipulated. It is not uncommon to adopt a ``safer'' approach of reducing  discrepancies with expectation by increasing systematics in the cases when there is no full confidence in the size of systematic uncertainties (in which case the most conservative assumption is used). 

As the result, this leads to a ``semi-blinded'' approach in a soft understanding of the blinding strategy, 
i.e. a blinding element is used initially, but further post-blinding manipulations with data are still allowed. 
This is particularly relevant for  the cases when there  are
no independent analysis teams involved in analysis.
A recent  example of the type B blinding 
can be found in \cite{Ablikim:2020hsk} where an observation of a near-threshold structure in the $K^{+}$ recoil-mass spectra in $e^+e^-$ collision was reported by the BES III  collaboration. 

%%%%%%%%%%%%%%%%%%%%%%%
\section{Type C}
%%%%%%%%%%%%%%%%%%%%%%%%%

In practice, a good theoretical understanding of signal regions in terms
of predictions may not be possible. 
The type C blinding deals with the following situations:

\begin{itemize}
    \item Theoretical predictions have significant uncertainties, i.e. larger than uncertainties expected for the signal region;
    
    \item  Monte Carlo simulations used for the description of background have significantly lower statistics than data;
    
    \item Control region in data  is not expected to catch all the kinematic details of the signal region. For example, it has a different physics menu or some reconstruction cuts. 
\end{itemize}

To overcome the above problems, a small fraction of "unblinded" signal region can be used (typically, this fraction is determined by looking at previously published low-statistics data).  As the result, generally, no strong requirement "not to look" at data can be imposed.

The type C blinding can be used as a guiding principle to perform some basic checks before looking at a signal region of data. 
Streaky speaking, the type C is a method to make ``an educational guess'' about background behavior in a signal region, 
but it cannot give the full confidence in our understanding of background (i.e. its shape and event rates).
None of the above studies at the pre-unblinding steps  guarantee that the background for signal region is sufficiently well understood at the level required for a proper blinding procedure type A or B.  Therefore, analysis team(s) should take a certain risk during opening the signal region, and should be prepared to see deviations in the signal region from the established background hypothesis.  Surely, such deviations do not need to be related to new physics. As the result, extensive cross checks have to be carried out with unblinded data to convince the community in  observation of genuine new physics. All such checks do not fall under the blinding principles since data can be manipulated one way or the other. 

Blinding  of the class C in searches 
can be found in \cite{Aad_2020,Aad:2016shx} and many other similar publications.

\section{Type D}

This technique does not assume blinding using quantitative estimates of shapes and normalizations of SM backgrounds.  
This type of blinding is appropriate when no well-understood theoretical predictions exist, 
nor a data control region. All object selections are standard and  there is no need to design complex phase space regions to enhance the signal-over-background ratio.

Generally, analyzers should have some qualitative 
expectations of how the SM background should look like, 
but they do not have precise quantitative 
predictions neither for SM  background nor for the BSM signal events.
For example, when searching for BSM signals 
in invariant masses (or jet masses), it is expected that the background is a smoothly falling distribution above the Sudakov peak, while signals can be seen as bell-shaped enhancements on top of smoothly falling data spectra\footnote{One can argue that the type D means ``no-blinding'', but we still prefer to call it as a variation  of the blinding technique since qualitative predictions are typically known from general kinematic arguments or previous low-statistic observations.}. Expectations for a smoothly falling background can be included in some analytic functions with unknown parameters.

For scenario D, extensive posterior checks are expected before claiming a discovery. Therefore, pre-unblinding preparation can be significantly reduced, or not used at all. The analyzers can look directly at the data using established performance selection cuts for all the objects used in the analysis. It is assumed that  no modifications of such selection criteria must be done. In this sense, analyzers ``blindly" follow the recommended object selections 
(jets, leptons and photons) provided by 
performance groups that are not directly involved in such searches.

Typical examples of the blinding D are searches in dijet invariant masses \cite{Sirunyan:2018xlo}, angular distributions derived from the rapidity of the two jets \cite{Aad_2016}, jet masses etc. (here we give only one reference per measurement type). 
In all such measurements, QCD predictions are not at the same level of precision as 
required for BSM searches in the signal region. 
However, qualitatively, we expect that the background shape is a falling function, while a signal has a bell-shaped form.  
Typically, the type D searches are combined with SM measurements. One striking example of the type D is an evidence for 
the top quarks \cite{Abe:1994xt} at the  Tevatron. The analyzers knew about possible signatures of top quarks, and made an effort to estimate SM background using Monte Carlo and control-region of data.
What comes out was an excess of events near 174 GeV above the estimated background. Multiple posterior tests could not reduce this excess. The observation of hadronic $W/Z$ decays in two-jet invariant masses
by the UA2 \cite{UA2:1990gao} also used an assumption on the approximate  shape of 
SM background, without any detailed knowledge of the SM predictions (and without "blinding").
Similarly, the observation of exotic structures in the $J/\Psi p$ channel \cite{Aaij:2015tga}, that can be interpreted as pentaquark, was made following the general knowledge of how this exotic state can decay, and what reconstruction steps should be undertaken to find it.

\section{Summary}

Although all the above types of blind analysis are expected to be well
applied to the real-world LHC studies, it is unlikely that the types A and  B 
are the best representation of high-precision
searches for BSM physics  at the LHC. 
The reason for this is following: it  is a rare case when there are several independent groups performing  the 
same analysis using different methods, thus ``post'' blinding checks with real data are going to happen anyway
in the case of unusual observations. The price to make mistakes in
claims of extraordinary discoveries is too high.
In addition, LHC searches in inclusive events (such as dijets, di-leptons) will deal with the level of statistical precision that is often  
significantly larger than theoretical uncertainties (or statistical precision of Monte Carlo simulations), thus blinding A/B cannot be used in such cases. 

As mentioned before, blinding A or B  is the most effective in the situations with several independent analysis teams
that are responsible for processing data and final analysis. For example, a technical  team ``blinds'' the signal region while the other teams  define  the analysis strategy based on the 
data blinded by the technical team.
If such a separation is impossible, the blinding strategy could be 
affected by  psychological effects that are not easy to overcome by small analysis teams with easy access to data since a   
signal region can be looked at (intentionally or unintentionally). 
The most common situation at the LHC is when the same analysis team performs 
many levels of data processing, including the reconstruction of  signal regions.

It is not unreasonable to think that, in the case of inclusive observables,
the type D approach, that does not  elevate blinding to the ``absolute necessity'', is the most sensible approach. It does not require a precise
understanding of theory nor SM backgrounds. However, it must heavily rely on recommendations for object reconstructions that  are typically developed by the teams that are not directly involved in searches. Another requirement is  to have a simple kinematic phase space that does not require complex selection cuts. For example, diphoton or dijet masses are typical examples because no a special selection is required to enhance signal regions. In this case,
the blinding means ``blindingly follow''  analysis recommendations of performance groups to reconstruct and identify objects, and build final observables for searches with a clear understanding of how unusual BSM events may look like. 

Even in more extreme, searches for unusual kinematic features in events where precise theoretical calculations are missing, can be  prioritized over other methods. Many major discoveries in the past, such as observation of $W$ \cite{Arnison:1983rp}, the discovery of gluon \cite{Brandelik:1979bd}, unusually higher rate of  diffracting events in $ep$ \cite{Derrick:1993xh},
observations of the top quarks \cite{Abe:1995hr,D0:1995jca} were done
without the blinding techniques A, B and C. 
Observations of unusual events as a byproduct of measurements with significant posterior checks to avoid ``false-positive'' is a fully justifiable path for new discoveries.
Many such studies can be  a part of SM measurements, or searches that use SM measurements in a combination  with qualitative assumptions on  how unusual BSM events should look like.

Directly looking at high-precision data when searching for unusual features, and performing  extensive posterior checks if such features are found, can be more appropriate and faster than ``semi-blinding'' methods (B, C). The latter methods may
significantly delay analyses while performing studies of  various phenomenological models, dealing low precision simulations, or developing systematic uncertainties on statistical limits for theoretical models even before seeing actual data. For example, when  it comes to searches of ``bumps'' in dijet masses above QCD background in inclusive events, our understanding of QCD processes  (which are dominant backgrounds for many searches) 
is at the level of a few percent \cite{Campbell:2013qaa} while typical searches for enhancements in dijet masses 
are performed  with a relative precision below  a few permille \cite{Aad:2019hjw}. 
Instrumental effects are also  larger than the precision with which data are probed
when looking for  new  physics in high-statistics LHC data. In such situations, blinding methods cannot reduce the risks of observations of spurious signals, while {\em posterior} checks have significantly larger value in reducing ``false-positives''.

The ''eureka moment'' is often the result of a
careful examination of data and explanation of unusual effects, 
rather than blinding strategies 
based on models that ``lock'' the attention of analyzers to a restrictive parameter domain of some narrowly-designed BSM physics. 
As argued before, even when using the blinding method B and C for  observing unusual features above a background level, a significant effort must be invested in exploring such new features, i.e.  by modifying selection cuts and by looking for possible systematic effects that may 
cause this feature.
Such checks do not fall into the paradigm of the strict ``blind'' strategy A, and the entire blinding procedure will be put into the doubt  and may even lose its merit.

It is advisable that analyzers agree about what type of blinding should be used prior to searches, and describe the type of blinding in final publications,  which may reduce confusion and possible misinterpretations by the readers.

\hspace{1cm}

{\it Acknowledgements.} I would like to thank S.R. Meehan for the discussion and comments. 
The submitted manuscript has been created by UChicago Argonne, LLC, Operator of Argonne National Laboratory (“Argonne”). Argonne, a U.S. 
Department of Energy Office of Science laboratory, is operated under Contract No. DE-AC02-06CH11357.  Argonne National Laboratory’s work was 
funded by the U.S. Department of Energy, Office of High Energy Physics under contract DE-AC02-06CH11357. 

%%%%%%%%%%%%%%%%%%%%%% references %%%%%%%%%%%%%%%%%%%%%%%%%%%%%%
%\section*{References}

\section*{References}
\bibliographystyle{elsarticle-num}
\def\bibname{\Large\bf References}
\bibliography{references}

\end{document}